\documentclass{amsproc}

\usepackage{amsmath}
\usepackage{amssymb}

\newcommand{\remove}[1]{}
\advance \topmargin by -\headheight
\advance \topmargin by -\headsep
\evensidemargin \oddsidemargin
\newtheorem{thm}{Theorem}      % A counter for Theorems etc
\newcommand{\BT}{\begin{thm}}   \newcommand{\ET}{\end{thm}}
\newtheorem{ithm}[thm]{Theorem (informal)}
\newcommand{\BIT}{\begin{ithm}}   \newcommand{\EIT}{\end{ithm}}
\newtheorem{dfn}{Definition}      %
\newcommand{\BD}{\begin{dfn}}   \newcommand{\ED}{\end{dfn}}
\newtheorem{corr}{Corollary}      %
\newcommand{\BCR}{\begin{corr}} \newcommand{\ECR}{\end{corr}}
\newtheorem{constr}[thm]{Construction}
\newcommand{\BCT}{\begin{constr}} \newcommand{\ECT}{\end{constr}}
\newtheorem{prop}{Proposition}
\newcommand{\BP}{\begin{prop}}   \newcommand{\EP}{\end{prop}}
\newtheorem{lemm}{Lemma}   % A counter for Lemmas etc
\newcommand{\BL}{\begin{lemm}}   \newcommand{\EL}{\end{lemm}}
\newtheorem{clm}{Claim}            %
\newcommand{\BCM}{\begin{clm}}   \newcommand{\ECM}{\end{clm}}
\newtheorem{sclm}{Sub-Claim}            %
\newcommand{\BSCM}{\begin{sclm}}   \newcommand{\ESCM}{\end{sclm}}
\newtheorem{fact}{Fact}            %
\newcommand{\BF}{\begin{fact}}   \newcommand{\EF}{\end{fact}}

\def\blackslug
{\hbox{\hskip 1pt\vrule width 8pt height 8pt depth 1.5pt\hskip 1pt}}
\def\qed{\quad\blackslug\lower 8.5pt\null\par}
\def\qqed{$\Box$}
\newenvironment{gproof}{\noindent{\bf Proof Sketch:~~}}{\qed}
\newcommand{\BPF}{\begin{gproof}} \newcommand {\EPF}{\end{gproof}}
\newenvironment{fproof}{\noindent{\bf Proof:~~}}{\qed}
\newcommand{\BPRF}{\begin{fproof}} \newcommand {\EPRF}{\end{fproof}}
\newenvironment{cproof}{\noindent{\bf Proof:~~}}{\qqed}
\newcommand{\BPRC}{\begin{cproof}} \newcommand {\EPRC}{\end{cproof}}
\newenvironment{hproof}{\noindent{\bf Proof~}}{\qed}
\newcommand{\BPR}{\begin{hproof}} \newcommand {\EPR}{\end{hproof}}

 \newenvironment{proofo}{\noindent{\bf Proof omitted.~~}}{}
\newcommand{\BPRO}{\begin{proofo}}\newcommand {\EPRO}{\end{proofo}
\vspace{0.2in}\\}

\newcommand{\BI}{\begin{itemize}}
\newcommand{\EI}{\end{itemize}}
\newcommand{\BE}{\begin{enumerate}}
\newcommand{\EE}{\end{enumerate}}

\begin{document}

\title{ Lower bounds for Arrangement-based Range-Free Localization in Sensor Networks}
\author{ Sandeep Gupta and Chinya V. Ravishankar}
\address{ Department of Computer Science Engineering \\
	University of California Riverside \\
	\{	sandeep,  ravi\}@cs.ucr.edu
}

%\authorblockA{Department of Computer Science Engineering \\
%	University of California Riverside \\%
%	\{	sandeep,  ravi\}@cs.ucr.edu
%}
%}
\begin{abstract}
Colander are location aware entities that collaborate to 
determine approximate location of mobile or static objects 
when beacons from an object are received by all colanders that are
within its distance $R$. This model, referred to as arrangement-based localization, does not require distance  estimation
between entities, which has been shown to be highly  erroneous in practice.
Colander are applicable in localization in sensor networks and  tracking
of mobile objects.

A set $S \subset {\mathbb R}^2$ is an $(R,\epsilon)$-colander if
by placing receivers at the points of $S$, a wireless device  with
transmission radius $R$ can be localized to within a circle of radius
$\epsilon$.  We present tight upper and lower bounds on the size of
$(R,\epsilon)$-colanders. We measure the expected size of colanders that
will form $(R,  \epsilon)$-colanders if they distributed uniformly over
the plane.

\end{abstract}		
\maketitle

\section{Introduction}
In several applications such as sensor networks, mobile ad-hoc networks
and location based services, one of the core issues is to obtain the
location of the wireless devices.

Colander are location aware entities
that collaborate to determine approximate location of mobile or static
objects when beacons from an object are received by all colanders that are
within its distance $R$. This model does not require distance  estimation
between entities, which is the  basis for most previous localization work.

In this work we analyze colanders as a tool to study the complexity of
localization in sensor networks.  We find lower and upper bounds for
the number of colander(" localizing devices") required in such a scheme.

Colander are applicable in localization in sensor networks and  tracking
of mobile objects. In this model, the moving object is required to echo
beacons at regular duration in order for the colanders to be able to
track it.

In~\cite{pattem}, authors propose a localization scheme where a moving
beacon echoes its coordinates; the sensors use this information to localize
their location.   Our results here derive the minimum number of
coordinates information required to be transmitted by the moving beacon
to achieve the  given  accuracy in localization.

Stupp and Sidi~\cite{expUnc} analyzed the accuracy of this localization
scheme in terms of expected area of uncertainty of position per sensor
when the anchors are distributed randomly on the plane. Also to make the
analysis easier the assume that the communication region of any node  is
a square for side length $R$ and that the distance is metric $L_{\infty}$
as oppose to $L_2$.

The problem of localization has previously been studied under different
settings. The problem addressed in this work is  similar in spirit to
that in~\cite{Demaine}. Authors in~\cite{Demaine} looked at the problem
of placing ''reflector'' on the walls of convex region so that any robot
equipped with "lasers" can always determine its current location.

\section{Range Free Localization}
Localization in sensor network is a widely studied problem.  These works
can be broadly classified into two categories: range-based and range-free.  Range-based technique rely on point-to-point distance estimates
or angle estimates to infer location. On the other hand, range-free
techniques do not use such estimations.

Because point-to-point distance estimates are highly erroneous,
range-free localization is being sought as a popular cost-effective
alternative to range based localization.

In~\cite{APIT} authors proposed Approximate Point in Triangle Test
(APIT) in which given three beacon nodes any sensor node can determine (in
range-free manner) if it lies inside the triangle. In there localization
scheme each sensor node performs numerous APT test with different
combination of audible anchor nodes and infers its location as the
center of gravity of the intersection area all the triangle in which it
lies in.  In~\cite{ROC} authors present a similar strategy but the test
uses concentric circles.

Liu and Wu~\cite{Wu} empirically study the effect of deployment of anchor
nodes on range free localization.

Although few papers exists for theoretical analysis of range based localization, no such study has been made for the range-free counterpart.
Our's is the first step toward a theoretical analysis for range free
localization, albeit the localized scheme here is  much simpler than
the recently proposed schemes. We present tight lower bounds on the
number of anchors required to achieve a minimum threshold on  uncertainty in localization.

\section{Problem}
For a point $p$ and $R \geq 0$, Let $B(p, R)$ be the closed ball of radius $R$ centered at $p$. Let $B(P, R) = \bigcup_{p\in P} B(p,R)$.
We call a set $S\subset \mathbb R^2$ an {\it $(R, \epsilon)$-colander} for a set $D \subset \mathbb R^2$ if for all points $p, p' \in D$ for which $B(p,R)\bigcap S = B(p',R) \bigcap S$, we have $||p-p'|| \leq \epsilon$. A more general definition of colander is got by asking that for a more general shape $W(p)$, $W(p) \bigcap S = W(p') \bigcap S$ implies $|| p - p'|| \leq \epsilon$.

We focus on the case $D$ being $[0,a]^2$ and $R < a/2$.
For $\epsilon > 0$, we ask what the size of the smallest $(R,\epsilon)$-colander for $D$ is.

\section{VC-dimension}
Our lower bounds use the results from
VC-dimension~\cite{VC} and $\epsilon$-nets~\cite{haus}. The
VC-dimension was first introduced by Vapnik and Chervonenkis,
and is a core concept in  statistical learning theory~\cite{learning}. 
 A {\it range space\/} is a pair
$(X,\mathcal{R})$, where $X$ is a set and $\mathcal{R}$ denotes the family of
subsets
of $X$ called {\it ranges\/}. The Vapnik-Chervonenkis or {\it VC-dimension\/},
of a range space $(X,\mathcal{R})$ is the cardinality of the largest subset of
$A\subseteq X$ for which \{$A \cap R\,|\, R \in \mathcal{R}$\} is the power
set of $A$. As an example consider range space $S = (X,\mathcal{C})$, where
$X$ is a set of points in the plane and $C$ are subsets lying on
circles. The VC-dimension of $S$ is 3, since if $Y\subseteq X$ is of size 3
or less, then for every subset $Y' \subseteq Y$, there exists a circle $C$ such
that $X  \cap C = Y'$, but this guarantee can not be made for larger
$Y$~\cite{bookjiri}.

\section{Upper Bound}
To motivate our construction, we first construct an infinite $(R, 0)$-colander of $0$ measure. Let $I = \{2xR: x \in \mathbb Z, 0 \leq x \leq 1 + 1/2R\}$. Let $S_v = I \times [0,1]$, $S_h = [0,1] \times I$ and $S = S_v \bigcup S_h$ (a square mesh with holes of size $2R \times 2R$).
We claim that $S$ is an $(R, 0)$-colander.
To see this, take a point $p = (x,y)$. If $x \neq (2k+1)R$ for some integer $k$, then $B(p, R) \bigcap S_v$ is a line segment, with $p$ lying on its perpendicular bisector (hence determining $y$). If $x = (2k+1)R$, then $B(p,R) \bigcap S_v$ is a set of $2$ points, with $p$ being their midpoint. Similarly, $B(p,R) \bigcap S_h$ would determine where $x$.

A slight generalization of this trick will give a $(R,\epsilon)$-colander. Define $J = \{y\epsilon/\sqrt{2}: y \in \mathbb Z, 0 \leq y \leq \sqrt{2}/\epsilon + 1\}$ and $K = \{xR: x \in \mathbb Z, 0 \leq x \leq 1/R + 1\}$. Let $S_v = K \times J$, $S_h = J \times K$ and $S = S_h \bigcup S_v$. To see that $S$ is an $(R, \epsilon)$-colander, take $p = (x,y)$. Note that any $B(p,R)$ will intersect $S_v$ in discrete points of some line segments, from which $y$ can be inferred up to distance $\epsilon/\sqrt{2}$. Similarly, $x$ can be inferred from $B(p,R) \bigcap S_h$ up to distance $\epsilon /\sqrt{2}$. Thus, $p$ can be determined up to distance $\epsilon$.

A similar construction can be given for $[0,a]^2$. In this construction, $|S|/a^2$ is $O(\frac{1}{R \epsilon})$.

\section{Lower Bound}

Given $S$, an $(R, \epsilon)$-colander for $[0, a]^2$, we wish to give a lower bound for $|S|$.

Define $S(p)$ to be $B(p,R) \bigcap S$. Define $p \equiv p'$ if $S(p) = S(p')$. Clearly $\equiv$ is an equivalence relation. Let us call these equivalence classes $\epsilon$-confusable regions. Any two points in an $\epsilon$-confusable region are at most distance $\epsilon$ apart (by definition of colander).

We will first show that a set of $n$ points can produce only $O(n^2)$ $\epsilon$-confusable regions. This alone will not give the tightest lower bound. Then, with a small partitioning trick, we will arrive at a lower bound of $\Omega(\frac{1}{R \epsilon})$, matching the upper bound.

Along the way, we will also show how lower bounds for colanders for other shapes can be computed.

\subsection{Duality}
Consider the mapping $\mathcal T_R$ taking $p$ to $B(p, R)$ and $B(p', R)$ to $p'$. Then $p \in B(p',R) \Leftrightarrow \mathcal T_R(B(p',R)) \in \mathcal T_R(p)$. Then $f$ is a duality between points and balls, with $\in$ replaced with $\ni$.

In particular, the following statements are equivalent:

$$ X \subset \bigcap_{p\in P} B(p,R)$$

$$ P \subset \bigcap_{x\in X} B(x,R) $$

We also define a more general version of duality that is applicable to any kind of shape. Given any $f: \mathbb R^2 \rightarrow \mathcal P(\mathbb R^2)$, define $f^* : \mathbb R^2 \rightarrow \mathcal P(\mathbb R^2)$, the dual of $f$, by $f^*(x) = \{y : x \in f(y)\}$. In the more general mapping, $p$ gets taken to $f(p)$ and $f^*(p)$ gets taken to $p$. This gives a similar duality.

\subsection{VC Dimension}

A range space $(X, \mathcal R)$ is a set $X$ along with a family of its subsets, $\mathcal R$. For $A \subset X$, define $P_{\mathcal R}(A) = |\{A \bigcap r: r \in \mathcal R\}|$. We say that $A$ is shattered by $(X, \mathcal R)$ if $P_{\mathcal R}(A) = 2^{|A|}$. The {\em VC Dimension}~\cite{VC} of a range space is the largest $n$ for which their exists a set of cardinality $n$ which is shattered by it.

Define $g(n,d) = \sum_{i=0}^d {n \choose i}$.
\begin{lemm}
(Sauer Lemma~\cite{Alon92}) For a range space $(X,R)$ with VC dimension $d$, for any $A \subset X$ with $|A| = n$, we have $P_{\mathcal R}(A) \leq g(n,d)$. 
\end{lemm}

\begin{lemm}
(Folklore) Let $\mathcal C$ be the family of closed discs in the plane. The VC dimension of $(\mathbb R^2, \mathcal C)$ is 3.
\end{lemm}
\BPRF
Clearly, any set of three non-collinear points can be shattered by $\mathcal C$. We will now show that any set $A$ of four points cannot be shattered. If their convex hull consists of three points, then any circle that contains those three is forced to contain the fourth.
So, assume that all the elements of $A$ are part of the convex hull of $A$. Let them be $a, b, c, d$ in that order. Let $e$ be the intersection of $ac$ and $bd$. Then, if $A$ could be shattered, there exists a circle $C_1$ containing only $\{a, c\}$ (and hence $e$) and another $C_2$ containing only $\{b, d\}$ (and hence $e$). Then $C_1$ and $C_2$ together would divide the plane into $5$ bounded regions (one for each of $a, b, c, d, e$), while two intersecting circles create only $3$ such regions. Thus, no set of $4$ points can be shattered by $\mathcal C$.
\EPRF

The next lemma, although trivial, captures what is really going on in colanders and relates it to circle arrangements on the plane.

\begin{lemm}
\label{dua}
Given a set $T$ of points on the plane. The number of different $B(p,R) \bigcap T$ as $p$ varies over all the points on the plane is no more than the number of regions produced by the following set of disks: $\{B(t,R): t \in T\}$.
\end{lemm}
\BPRF
Follows directly from the duality. From first principles, the key observation is that if $B(p,R) \bigcap T \neq B(q,R) \bigcap T$, then there is a circle $B(t,R), t \in T$ that contains $p$ and not $q$ or contains $q$ and not $p$, and hence $p$ and $q$ are in different regions.
\EPRF

\begin{lemm}
The number of regions created by $n$ equal circles on the plane $\leq g(n,3)$.
\end{lemm}
\BPRF
For any set $A$ of $n$ points, $P_\mathcal{C}(A) \leq g(n,3)$. By lemma \ref{dua}, the result follows.
\EPRF

Although the above result is weaker than the best, it easily generalizes to the following with the more general version of duality:

\begin{lemm}
Consider a convex set $S$. Let the family of its translates on the plane be $\mathcal S$. Let the VC dimension of $(\mathbb R^2, \mathcal S)$ be $d$. The number of faces in an arrangement of $n$ $S$'s on the on the plane is $O(n^d)$.
\end{lemm}

This, in turn, can give a lower bound on the size of colanders (using arguments similar to what we will see in Theorem \ref{th:lb}) for other shapes ($\Omega(1/R^{2-2/d}\epsilon^{2/d}$, where $d$ is the VC dimension). For the case of circles, it will give $\Omega(1/R^{4/3}\epsilon^{2/3})$, which does not match the upper bound. Fortunately, the following result for the special case of circles, does better than the VC-dimension machinery we built up and lets us get a tight bound.

\begin{thm}
(Folklore) $n$ equal circles on the plane divide it into at most $O(n^2)$ regions~\cite{mathworld}. 
\end{thm}

\begin{thm}
Any $(R,\epsilon)$-colander for a region of area $A$ must have at least $\sqrt{A/{\epsilon^2}}$ points.
\end{thm}
\BPRF
Every $\epsilon$-confusable region fits within a circle of diameter $2\epsilon$, and so there must be at least $A/\pi\epsilon^2$ $\epsilon$-confusable regions. However, $n$ circles on the plane can produce only $n^2$ regions, and so by lemma \ref{dua}, the result follows.
\EPRF

Note that this gives a weak lower bound of $|S|/a^2 \geq \frac{1}{a\epsilon}$ (that reduces as the area increases: weird?). However, we will use this to bootstrap our next lower bound in Theorem \ref{th:lb}.

\begin{lemm}
Let $S$ be a $(R,\epsilon)$-colander for a region $D$, and let $p$ be a point with $B(p,R) \subset D$. Then $S \bigcap B(p,2R)$ is a $(R,\epsilon)$-colander for region $B(p,R)$.
\end{lemm}
\BPRF
For any point $q \in B(p,R)$, $B(q,R) \subset B(p,2R)$, and so $B(q,R) \bigcap (S \bigcap B(p,2R)) = B(q,R) \bigcap S$, and hence $q$ can be determined upto distance $\epsilon$, by the colander property of $S$.
\EPRF

\begin{thm}
\label{th:lb}
If $S$ is a $(R,\epsilon)$-colander for $[0,a]^2$, then $|S|/a^2 \geq \Omega(\frac{1}{R\epsilon})$
\end{thm}
\BPRF
Partition $[0,a]^2$ into $\frac{a^2}{16R^2} - o(1/R^2)$ square regions, each of size $4R \times 4R$. Consider one of the squares with its center $p$. Now, $S \bigcap B(p, 2R)$ is a $(R, \epsilon)$-colander for $B(p,R)$, and so there must be at least $R^2/\epsilon^2$ $\epsilon$-confusable regions, and hence at least $R/\epsilon$ points in $S \bigcap B(p,2R)$. Now, over all choices of $p$ (as center of a square), all the $B(p,2R)$ are disjoint, and hence $|S| \geq \sum_p |S\bigcap B(p,2R)| \geq \frac{a^2}{16R^2} \frac{R}{\epsilon} = \frac{a^2}{16R\epsilon}$.
\EPRF

\subsection{Colanders for Uniform Distribution}
In this section we derive the expected number of colanders
to be distributed uniformly in a plane to guarantee at $(R,
\epsilon)$-localization.

Its reasonable to assume that under uniform distribution all the
subelements cover same area, at least in expectation. We are forced to
make this assumption as the formula for computing the area of a subelement
quite involved, even for subelements formed with 3 circles.

Let $r$ colanders be distributed uniformly in  unit area, then the
expected number of subelements is $r^2$. For $(R, \epsilon)$ localization,
	$1/r^2 = 1/\epsilon$. Hence $r = \sqrt \epsilon$.

\section{Open Questions}

One question that we failed to resolve was the colander property of a uniform grid. That is, what is the smallest $\epsilon$ for which the set $\{(a\delta, b\delta): a,b \in \mathbb Z, 0 \leq a,b \leq 1/\delta + 1\}$ is a $(R,\epsilon)$-colander. It appears to have some deep connections with number theory, in particular, the distribution of numbers representable as the sum of two squares. Other things that could be sought are better lower and upper bounds for colanders for different shapes.
\section{Conclusion}

Following question is still unanswered regarding tracking of moving object: For a given speed of the moving object

%\input{conclusion}
%\bibliography{collated}

\end{document}